\documentclass[aps,prl,twocolumn,groupedaddress]{revtex4}
\usepackage{graphicx}

\begin{document}

%\preprint{}

\title{Realistic network growth using only local information: \\
From random to scale-free and beyond}


\author{David M.D. Smith}
\thanks{Corresponding author (d.smith3@physics.ox.ac.uk).}
\author{Chiu Fan \surname{Lee}}
\author{Neil F. Johnson}
%\thanks{}
\affiliation{Physics Department, Clarendon Laboratory, Oxford University, Oxford OX1 %%@
3PU, U.K.}


\date{\today}

\begin{abstract}
We introduce a simple one-parameter network growth algorithm which is able to reproduce a wide variety of realistic network structures {\em without} having to invoke any global information about node degrees such as preferential-attachment probabilities. Scale-free networks arise at the transition point between quasi-random and quasi-ordered networks. We provide a detailed formalism which accurately describes the entire network range, including this critical point. Our formalism is built around a statistical description of the inter-node linkages, as opposed to the single-node degrees, and can be applied to any real-world network -- in particular, those where node-node degree correlations might be important. 
\end{abstract}

%\keywords{}

\maketitle


Networks -- in particular, large networks with many nodes and links -- are attracting increasing attention from researchers in the fields of physics and biology through to sociology, informatics and medicine \cite{random,Newman,Barabasi,pakming,Holme,Leary,Krapivsky,Saramaki,Evans,Batagelj98,Callaway,Vazquez}. Many social and informational networks, including the World Wide Web, exhibit scale-free behaviour while others lie closer to fully-random or fully-ordered \cite{random,Newman}. Growth mechanisms have been proposed for a wide variety of such networks -- however, many of these growth mechanisms require some form of global knowledge of the vertex (i.e. node) degree. In particular, the famous preferential-attachment model of Barab\'asi and Albert \cite{Barabasi} requires knowledge of each node's degree in order to evaluate the corresponding attachment probability. Holme and Kim \cite{Holme} and Leary {\it et al} \cite{Leary} %%@
have subsequently considered a modification of this preferential rule by ranking existing nodes according %%@
to degree, and associating
corresponding probabilities of connection with the new node accordingly. Their approach led to over- and %%@
under-skewed distributions for the node degree, as compared to the  
power-law distribution characteristic of scale-free networks. 
However in both the original and modified preferential-attachment algorithms,
information is required about the
degree -- and hence preferential-attachment probability -- of every node in the network, prior to adding a new node or link. 

Although preferential attachment models can yield similar distributions to real-world systems, this requirement that the next node or link  `knows' about the degree of every node in a large network, makes the mechanism microscopically unrealistic for many real-world networks. In particular, most biological and social networks are too large for such information to be accessible. Motivated by this shortcoming, the present paper discusses a simple, one-parameter algorithm which can reproduce  under-skewed, over-skewed and  
scale-free (i.e. power-law) networks {\em without} global knowledge of the node degrees -- see Figs. 1 and 2. It therefore provides an alternative, and arguably more realistic, microscopic mechanism for a wide range of biological and social networks, whose growth had previously been explained using the Barab\'asi-Albert preferential--attachment model \cite{Barabasi}. As a by-product of our analysis, we also provide a new formalism to describe general network growth %%@
dynamics {\em including} node-node linkage correlations. When applied to our growth algorithm, this `link-space' formalism allows us to identify the transition point at which over-skewed node-degree distributions switch to under-skewed distributions. It is at this transition point, and only at this point, that we find scale-free networks emerge. %%@
Similar notions of transitive linking have been studied elsewhere \cite{Krapivsky} -- in particular, Saramaki %%@
and 
Kaski \cite{Saramaki} and 
Evans \cite{Evans} have recently presented some fascinating results using random-walkers to decide node attachment in 
a network-growth algorithm. However, these analyses were all mean-field in nature and %%@
hence were not able to accurately describe the node-node linkage correlations which can be a crucial feature of real-world networks, and are a crucial feature of the one-parameter networks discussed here.

\begin{figure}
\includegraphics[width=0.15\textwidth]{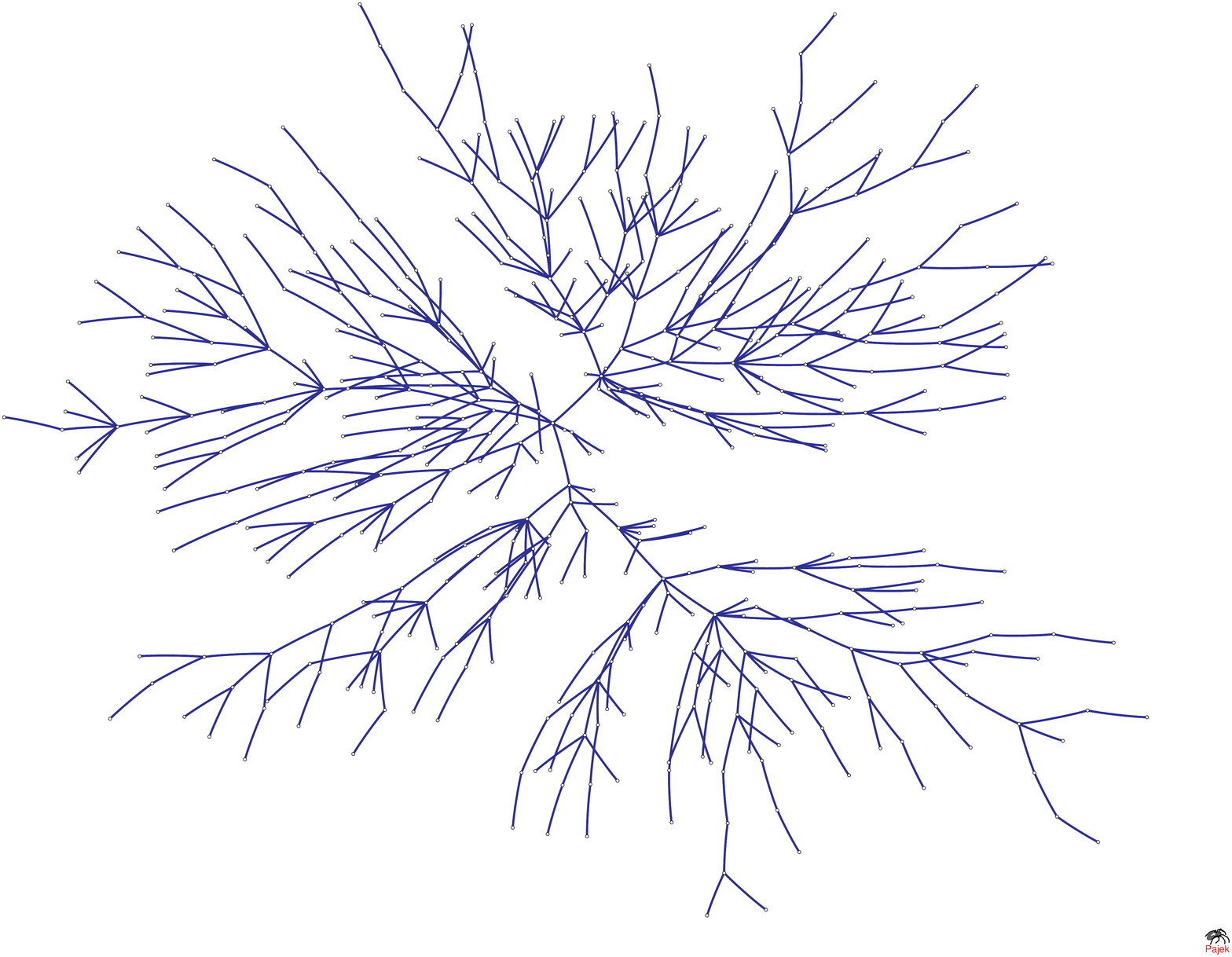}
\includegraphics[width=0.15\textwidth]{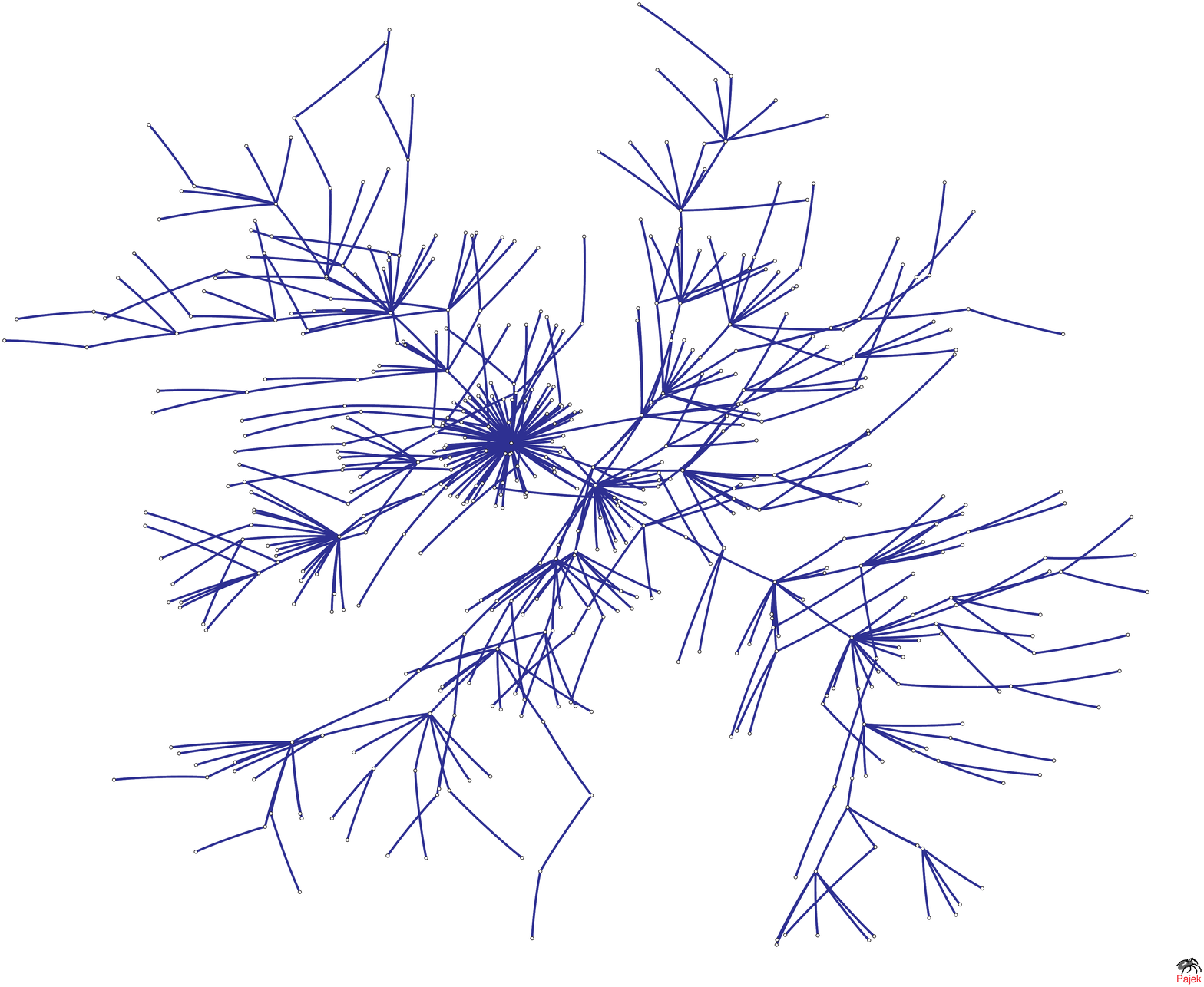}
\includegraphics[width=0.15\textwidth]{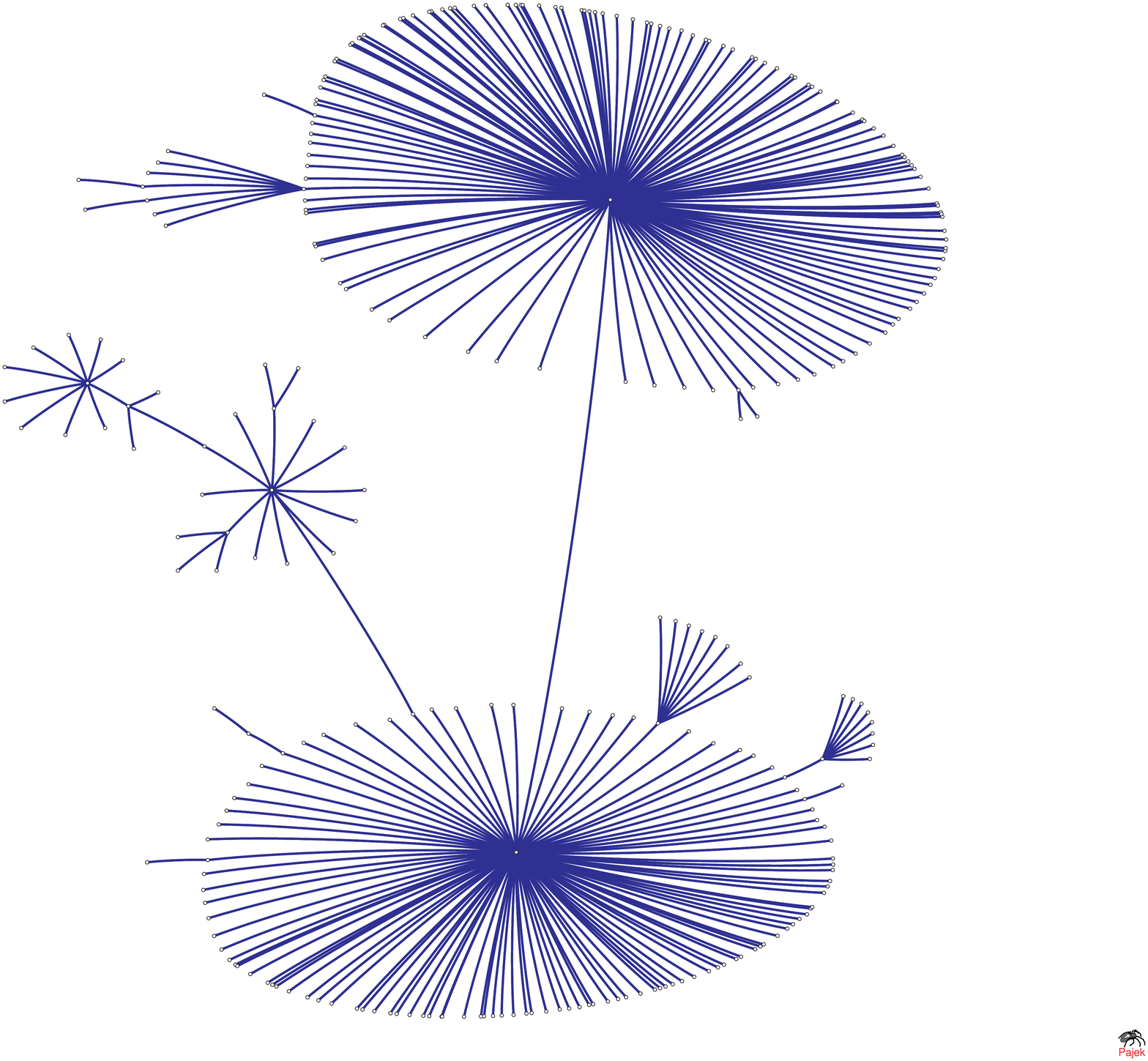}
\caption{ \label{fig:exNN} (Color Online) Networks generated by our one-parameter, local information growth algorithm with $a =1$ (left),  $a=0.2$ (center), $a=0$ (right).}
\end{figure}

\begin{figure}
%\begin{center}
\includegraphics[width=0.5\textwidth]{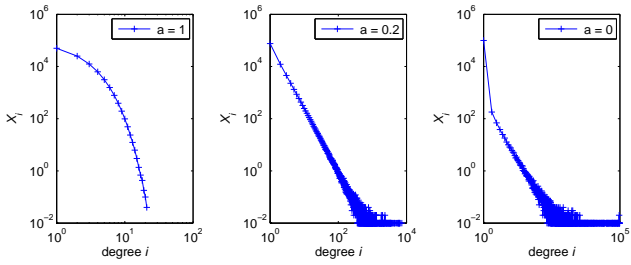}
%\end{center}
\caption{ \label{fig:results} (Color Online) Degree distributions for the networks in Fig. 1, grown %%@
to 100,000 nodes and ensemble-averaged over  
100 networks per $a$-value. Initial seed comprised two nodes and one link.}
\end{figure}

We first introduce our general formalism. We then use it to discuss various existing growth algorithms and our one-parameter growth algorithm, together on the same footing. For a growing network in which one node and undirected link are added at timestep $t$, we can write the %%@
master equation for $X_i$ (the number of nodes of degree $i$) in terms of the probabilities $\{\Theta_j\}$  for attaching to a
node of degree $j$:
\begin{equation}
X_i \mid_{t+1} = X_i \mid_t +\Theta_{i-1}\mid_t-\Theta_{i}\mid_t,\ \ \ i>1\ \ .
\label{eqn:nodeXmaster}
\end{equation}
The second term describes the probability of the new node attaching to an existing %%@
node of degree $i-1$ thereby making it a  
degree $i$ node. The third term on the right hand side describes the new node %%@
attaching to a node of degree $i$, thereby decreasing $X_i$.
For $i=1$, the second term is unity as a new node of degree one is added to the existing network each timestep. If we assume steady-state growth, then $X_i = c_i t$ and hence
the fraction of nodes of degree $i$ is constant ($\frac{dX_i}{dt}=c_i$). To facilitate the comparison with later expressions, we write the solution in the following form:
\begin{equation}
c_i = \frac{\Theta_{i-1}}{1+\frac{\Theta_i}{c_i}}
\label{eqn:nodespacemaster}
\end{equation} 
The notation `$\mid_{t}$' has been dropped to indicate steady-state.
So far we have said nothing about the attachment mechanism, and have made
the easily geneneralizable restriction that only one node and undirected link is being added per timestep.
We now follow a similar analysis, but retain the node-node linkage correlations that are inherent in many real-world systems \cite{Callaway,Krapivsky}.
Consider any link in a general network -- we can describe it by the degrees of the two nodes 
that it connects. 
Hence we can construct a matrix $\mathbf{L}$ such that the element
$L_{i,j}$ describes the  number of links from nodes of degree $i$ to nodes of degree $j$.
For undirected networks, the summation over all elements $L_{i,j}$ is equal to
twice the total number of links in the network. This matrix represents a surface %%@
describing the first-order correlations  
between the node degrees -- we refer to this as the {\em link space}.

In order to write the master equations for the evolution of the network in this link %%@
space, we must first understand the impact of adding a new link.
The probability of selecting any node of degree $i-1$ is given by the attachment %%@
kernel $\Theta_{i-1}$.
Suppose an $i-1$ node is selected -- 
the fraction of these
that are connected to nodes of degree $j$, is
$\frac{L_{i-1,j}\mid_t}{(i-1)X_{i-1}\mid_{t}}$. The term  
describing the increase 
in links from nodes of degree $i$ to nodes of degree $j$, through the %%@
attachment of the new node to a node  
of degree $i-1$, is given by:
\begin{equation}
\frac{\Theta_{i-1}\mid_{t} L_{i-1,j}\mid_t}{X_{i-1}\mid_{t}}\ \ \ .
\end{equation}
Since each link has two ends, the value $L_{i,j}$ can increase by %%@
connection to an $(i-1)$-degree node which is in turn connected to a $j$-degree node, or by %%@
connection to an $(j-1)$-degree node which is in turn connected to an $i$-degree node.
 
\begin{figure}
\includegraphics[width=0.45\textwidth]{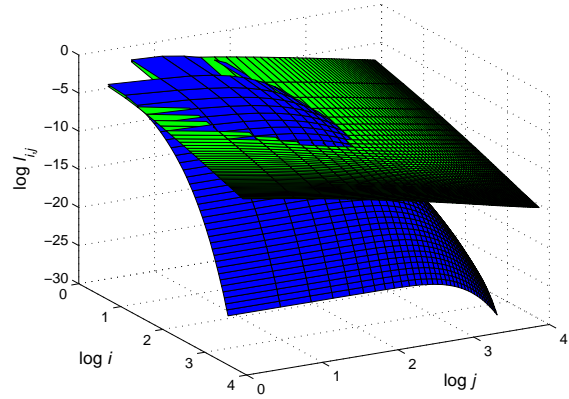}
\caption{ \label{fig:logloglogls} (Color Online) Representation of the link-space matrix with elements $l_{i,j}$, in logarithmic coordinates. The values shown are the steady-state solutions for the random attachment (curved surface) and preferential attachment (flatter surface) algorithms. The value $l_{1,1}$ is zero for networks consisting of a single component.}
\end{figure}

We now make a similar steady-state assumption to the `node space' example above, and obtain \cite{David}:
\begin{eqnarray}
l_{i,j}& =& \frac{\frac{\Theta_{i-1}}{c_{i-1}}l_{i-1,j} ~+~ %%@
\frac{\Theta_{j-1}}{c_{j-1}}l_{i,j-1}}{1~+ %%@
\frac{\Theta_{i}}{c_{i}}~+~\frac{\Theta_{j}}{c_{j}}  },\ \ \  ~~~i,j>1;  \nonumber\\
l_{1,j}& =& \frac{\frac{\Theta_{j-1}}{c_{j-1}}l_{1,j-1}~+~\Theta_{j-1}}{1~+ %%@
\frac{\Theta_{1}}{c_{1}}~+~\frac{\Theta_{j}}{c_{j}} }, \ \ \  ~~~~~~~~j>1;
\label{eqn:linkspacemaster}
\end{eqnarray}
where $l_{i,j}$ denotes the fraction of links that connect an $i$-degree node to a %%@
$j$-degree node.  It turns out that $l_{1,1}$ is zero for all the growth algorithms discussed in this paper, since the networks that they generate comprise a single component. %%@
The notation $\mid_{t}$ has been dropped as before, to indicate the steady state.
The fraction of $i$-degree nodes is:
\begin{equation}\label{eqn:ci}
c_i = \frac{\sum_k l_{i,k}}{i} \ \ , \ \
X_i = \frac{\sum_k L_{i,k}}{i} \ .
\end{equation}
Hence the degree distribution is retrievable from the link-space matrix. 
To illustrate use of the formalism, consider first a random-attachment model in which the existing node to which the new node is to be %%@
connected, is chosen randomly. The attachment probability is $\Theta_i~=~c_i$.
Substituting into Eq.~\ref{eqn:nodespacemaster}, we obtain the recurrence %%@
relation $c_{i+1}= \frac{c_i}{2}$ which yields the familiar $2^{-i}$ distribution of node degree.
Substituting into the link-space master equation (Eq.~\ref{eqn:linkspacemaster}) yields the recurrence relations:
\begin{eqnarray}\label{eqn:ramaster2}
l_{i,j}&=& \frac{l_{i-1,j}+l_{i,j-1}}{3}, \ \ \ \ ~~~~i,j~>1;\nonumber\\
l_{1,j}&=& \frac{c_{j-1}+l_{1,j-1}}{3},\ \ \ \ ~~~~j>1;\nonumber\\
l_{1,1}&=& 0\ \ \ .
\end{eqnarray}
The exact solution for $l_{i,j}$ is
\[
\sum_{\alpha=2}^{j} 
\frac{{}^{(i-1+j-\alpha)}C_{(j-\alpha)}}{3^{(i+j-\alpha)}2^{(\alpha-1)}} 
 + \sum_{\alpha=2}^{i} %%@
\frac{{}^{(i-1+j-\alpha)}C_{(i-\alpha)}}{3^{(i+j-\alpha)}2^{(\alpha-1)}},\ \ \ \ ~~~i,j>1;
\]
\begin{equation}\label{eqn:raexact}
 \sum_{k=1}^{j-1}\frac{1}{3^k 2^{j-k}},\ \ \ \ ~~~i=1,j>1,
\end{equation}
where $C$ is 
the conventional combinatorial `choose'  
function.
 The surface generated is shown in Fig.~\ref{fig:logloglogls}. 

In the Barab\'asi-Albert preferential attachment algorithm \cite{Barabasi}, the attachment %%@
probability is proportional to the degree of the node in the existing 
network:
\begin{equation}
\Theta_i\mid_t = \frac{i X_i}{2(N-1)}  
\label{eqn:SFkernel}
\end{equation}
which in the steady-state limit of large $N$ (i.e. many nodes) can be well %%@
approximated by $\Theta_i \approx  \frac{i c_i}{2}$.
Substituting this expression into Eq.~\ref{eqn:nodespacemaster} yields the
familiar recurrence relation
\begin{equation}
c_{i}=\frac{(i-1)~ c_{i-1}}{2}-\frac{i~c_i}{2}
=\frac{i-1}{i+2}c_{i-1}
\end{equation} 
whose solution is:
\begin{eqnarray}
\label{eqn:NSpref}
c_{i}&=&\frac{4}{i(i+1)(i+2)}\ \ \ .
\end{eqnarray} 
Using the same substitution, the link-space master equations yield the recurrence relations:
\begin{eqnarray}\label{eqn:sfmaster3}
l_{i,j}&=&\frac{(i-1)l_{i-1,j}~+~(j-1)l_{i,j-1}}{2+i+j},\ \ \ ~~~i,j>1;\nonumber\\
l_{1,j}&=&\frac{(j-1)c_{j-1}~+~(j-1)l_{1,j-1}}{3+j},\ \ \ ~~~j>1\  \ .
\end{eqnarray}
The exact solution for $l_{i,j}$ is obtained by algebraic manipulation of the link-space matrix 
using the previously derived degree distribution, and is given by \cite{David}:
\begin{equation}
\begin{array}{l}
l_{i,j}= \frac{4(j-1)!(i-1)!}{(j+i+2)!}\Big\{G(i+1)+2G(i)-3G(i-1) \nonumber\\ %%@
+\frac{1}{2}\sum_{\alpha=1}^{i}(\alpha-1)(\alpha+6)\left[ %%@
G(i-\alpha)-G(i-\alpha-1)\right] \Big\}; \nonumber\\
G(x)~=~ \left\{\begin{array}{l r}
\frac{(j+x-1)!}{x!(j-1)!}&\textrm{for $x\ge0$}\\
0 & \textrm{for  $x<0$} \ . \end{array}\right. 
\end{array}
\end{equation}
The first two rows of this matrix, have the form:
\begin{eqnarray}
l_{1,j}& =& \frac{2(6+j)(j-1)}{j(j+1)(j+2)(j+3)};\ \ \ \ \nonumber \\
l_{2,j}&=&\frac{2j(j-1)(j+10)+48}{3j(j+1)(j+2)(j+3)(j+4)} \ .
\label{eqn:sfexact}
\end{eqnarray}
The above equations for $l_{i,j}$, even when invoked to low order in the iteration scheme, can accurately reproduce the Barab\'asi-Albert preferential attachment network. The corresponding surface is shown in Fig.~\ref{fig:logloglogls}. 

We now turn to our one-parameter network growth model, which involves attaching a single  
node at each 
timestep {\em without} prior knowledge of the existing network structure.
The algorithm goes as follows:
i) Pick a node $\kappa$ within the existing network at random.
ii) With probability $a$ make a link to that node. Otherwise
iii) pick any of the neighbours of $\kappa$ at random, and link to that node. Hence this algorithm resembles an object or `agent' making a short random-walk.
Figure 1 shows examples of the resulting networks, with Fig. 2 showing the corresponding degree distributions.
Interestingly, $a=0$ yields a graph that is dominated by hubs and spokes %%@
(i.e. extreme  
over-skewed) \cite{footnote1}
while $a = 1 $ yields the 
random-attachment graph. Intermediate values of $a$ yield networks which are neither too ordered nor too disordered. For $a\sim 0.2$, the 
algorithm generates networks whose degree-distribution closely resembles Barab\'asi-Albert preferential-attachment networks (see Figs.~1, 2 and 4).

We now develop master equations for the evolution
of this one-parameter network generated with only local information. We first establish the attachment %%@
probability kernel $\Theta_i$ for this algorithm, which in turn requires properly resolving %%@
the one-step random walk. 
The link-space formalism provides us with an expression for the probability $P'_i$ %%@
associated with performing a  
random walk of length one and arriving at a node of degree $i$. Note that this is 
different to arriving at a {\em specific} node of  
degree $i$ after a one-step walk, since here we consider the possibility of %%@
arriving at {\em any} of the nodes which happen to  
have degree $i$ at time $t$ \cite{David}:
\begin{eqnarray}\label{eqn:neighbour}
P'_i&=& \frac{X_1\mid_{t} L_{1,i}\mid_{t}}{N_t  X_1\mid_{t}}+\frac{X_2\mid_{t} %%@
L_{2,i}\mid_{t}}{2 N_t   
X_2\mid_{t}}+\frac{X_3\mid_{t} L_{3,i}\mid_{t}}{3 N_t  X_3\mid_{t}}+\ldots %%@
\nonumber\\
{} &=& \frac{1}{N_t}\sum_k \frac{L_{k,i}\mid_{t}}{k}
\end{eqnarray}
This can also be written \cite{David} as $
P'_i =   \frac{i X_i\mid_{t}}{N_t}\langle\frac{1}{k}\rangle_i$
where the average is performed over the neighbours of nodes with degree $i$. Note %%@
that this quantity does {\em not} replicate 
preferential attachment, in contrast to what is commonly thought ~\cite{Saramaki, Vazquez}.
Defining $\beta_i$ as
\begin{equation}
\beta_i \equiv \frac{1}{i~c_i}\langle \frac{1}{k}\rangle_i = \frac{\sum_k %%@
\frac{L_{i,k}}{k}}{\sum_k L_{i,k}}
=\frac{\sum_k \frac{l_{i,k}}{k}}{\sum_k l_{i,k}} \ ,
\label{eqn:beta}
\end{equation}
yields
\begin{equation}
\Theta_i= a c_i~+~(1-a)\beta_i i c_i
\label{eqn:xnnkernel}
\end{equation}
Substituting Eq.~\ref{eqn:xnnkernel} into Eq.~\ref{eqn:nodespacemaster}, we obtain %%@
the following for the steady-state node degree:
\begin{equation}
c_i = \frac{c_{i-1}\big(a+(1-a)\beta_{i-1}(i-1)\big)}{1+a+(1-a)\beta_i i}\ \ .
\end{equation}
Substituting into Eq.~\ref{eqn:linkspacemaster} yields the following recurrence relations:
\begin{tiny}
\[
l_{i,j}=\frac{l_{i-1,j}\big(a+(1-a)\beta_{i-1}(i-1)\big)+l_{i,j-1}\big(a+(1-a)\beta_{j-1}(j-1%%@
%%@
)\big)}
{1+2a+(1-a)(i\beta_i+j\beta_j)}  
\]
\[
l_{1,j}=\frac{l_{1,j-1}\big(a+(1-a)\beta_{j-1}(j-1)\big)+c_{j-1}\big(a+(1-a)\beta_{j-1}(j-1)%%@
\big)}
{1+2a+(1-a)(\beta_1+j\beta_j)} \ .
\]
\end{tiny}
The non-linear terms resulting from $\beta$ imply that an explicit closed expression for $l_{i,j}$ is difficult. We leave this as a future challenge -- but we stress that our formalism  can  be implemented in its non-stationary form numerically
(i.e. iteratively)
with very good efficiency \cite{David}, as demonstrated by the degree distributions in Fig.~4.

\begin{figure}
\includegraphics[width=0.45\textwidth]{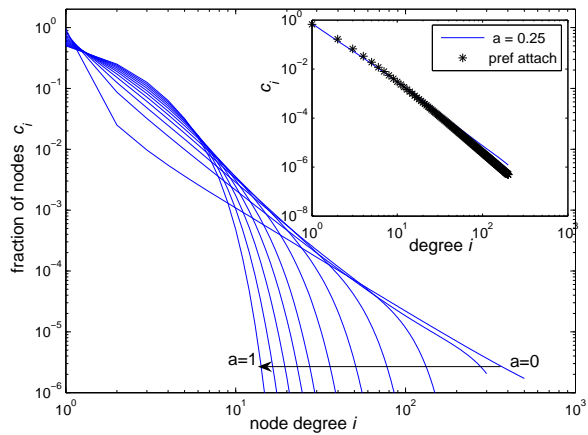}
\caption{ \label{fig:lsresults} (Color Online) Degree distributions generated using the link-space analysis of our one-step algorithm. Inset: results for $a=0.25$ reproduce the same distribution as for the Barab\'asi-Albert \cite{Barabasi} preferential-attachment algorithm.}
\end{figure}

Our numerical results suggests that for $a\sim0.2$, the degree distribution approaches %%@
scale-free. We now use the above formalism to deduce the %%@
critical value $a_c$ at which the node-degree distribution goes from over-skewed to %%@
under-skewed, and hence the value of $a$ at which a scale-free distribution arises. At $a_c$, we know that the attachment %%@
kernels for both our and the Barab\'asi-Albert model should be equal. Hence from
Eq. \ref{eqn:SFkernel} and Eq. \ref{eqn:xnnkernel}, we have
\begin{equation}
\frac{i}{2}= a_c~+~(1-a_c)\beta_i i 
\label{eqn:a}
\end{equation}
which for large $i$ yields $a_c=1-\frac{1}{2\beta_i}$.
We could proceed to use the exact solution of the link-space equations for the %%@
preferential-attachment algorithm, in order to infer $\beta_i$ in the high $i$ limit. However, since %%@
$\beta_i$ can be expanded in terms of $l_{i,j}$ as shown in Eq.~\ref{eqn:beta}, and %%@
$l_{i,j}$ decays very rapidly as $i,j$ become large, we can obtain a good approximation by %%@
only using the first two terms of Eq.~\ref{eqn:sfexact}. This gives $\beta\sim0.66$. Substituting into Eq.~\ref{eqn:a} then yields the critical value at which scale-free networks exist as $a_c=0.25$, in excellent agreement with the results shown in the inset plot of %%@
Fig.~4.

In conclusion, we have presented a simple, one-parameter algorithm for generating networks whose %%@
properties span from the exponential degree  
distribution of random 
attachment, through to an ordered `hub and spoke' situation (Fig. 1). A scale-free network turns out to be a special case, lying at a transition point in between the two limits. The growth algorithm utilizes only one 
simple parameter $a$, and requires no global information concerning node degree. As a by-product of the analysis of this model, we have managed to develop a new type of link-space formalism which can account for the node-node linkage correlations in real-world networks. Indeed, as will be shown elsewhere, the link-space formalism can be used to describe an even wider variety of networks than those emerging from the present growth model.


\begin{thebibliography}{99}
\bibitem{footnote1}
Suppose we seed the algorithm with an existing hub and spoke graph, and use $a = 0$. The  
probability that a new node will be linked to the hub tends to unity as the network %%@
grows. More generally, the distribution of vertex  
degrees will approach that of a hub and spoke network as the network grows. %%@
This type of network can be considered the  
most ordered in that the mean free path is minimized (i.e. it tends to two)
\end{thebibliography}
\end{document}